\documentclass[pra,aps,showpacs,onecolumn,twoside,superscriptaddress]{revtex4}

\usepackage{amsmath,amsfonts,amssymb,color,epsfig,graphics,graphicx,latexsym,revsymb,theorem,url,verbatim}

\newtheorem{definition}{Definition}
\newtheorem{proposition}[definition]{Proposition}
\newtheorem{lemma}[definition]{Lemma}

\newtheorem{theorem}[definition]{Theorem}
\newtheorem{corollary}[definition]{Corollary}
\newtheorem{conjecture}[definition]{Conjecture}

\newtheorem{remark}[definition]{Remark}
\newtheorem{example}[definition]{Example}
\newtheorem{question}[definition]{Question}

\def\squareforqed{\hbox{\rlap{$\sqcap$}$\sqcup$}}
\def\qed{\ifmmode\squareforqed\else{\unskip\nobreak\hfil
\penalty50\hskip1em\null\nobreak\hfil\squareforqed
\parfillskip=0pt\finalhyphendemerits=0\endgraf}\fi}
\def\endenv{\ifmmode\;\else{\unskip\nobreak\hfil
\penalty50\hskip1em\null\nobreak\hfil\;
\parfillskip=0pt\finalhyphendemerits=0\endgraf}\fi}
\newenvironment{proof}{\noindent \textbf{{Proof.~} }}{\qed}
\def\Dbar{\leavevmode\lower.6ex\hbox to 0pt
{\hskip-.23ex\accent"16\hss}D}
\makeatletter
\def\url@leostyle{%
  \@ifundefined{selectfont}{\def\UrlFont{\sf}}{\def\UrlFont{\small\ttfamily}}}
\makeatother
\def\bcj{\begin{conjecture}}
\def\ecj{\end{conjecture}}
\def\bcr{\begin{corollary}}
\def\ecr{\end{corollary}}
\def\bd{\begin{definition}}
\def\ed{\end{definition}}
\def\bea{\begin{eqnarray}}
\def\eea{\end{eqnarray}}
\def\bem{\begin{enumerate}}
\def\eem{\end{enumerate}}
\def\bex{\begin{example}}
\def\eex{\end{example}}
\def\bim{\begin{itemize}}
\def\eim{\end{itemize}}
\def\bl{\begin{lemma}}
\def\el{\end{lemma}}
\def\bpf{\begin{proof}}
\def\epf{\end{proof}}
\def\bpp{\begin{proposition}}
\def\epp{\end{proposition}}
\def\bqu{\begin{question}}
\def\equ{\end{question}}
\def\br{\begin{remark}}
\def\er{\end{remark}}
\def\bt{\begin{theorem}}
\def\et{\end{theorem}}

\def\btb{\begin{tabular}}
\def\etb{\end{tabular}}

\newcommand{\nc}{\newcommand}

\def\r{\rho}

\def\ps{\psi}

\def\G{\Gamma}

\def\Ps{\Psi}

 \nc{\bA}{{\bf A}} \nc{\bB}{{\bf B}} \nc{\bC}{{\bf C}}
 \nc{\bD}{{\bf D}} \nc{\bE}{{\bf E}} \nc{\bF}{{\bf F}}
 \nc{\bG}{{\bf G}} \nc{\bH}{{\bf H}} \nc{\bI}{{\bf I}}
 \nc{\bJ}{{\bf J}} \nc{\bK}{{\bf K}} \nc{\bL}{{\bf L}}
 \nc{\bM}{{\bf M}} \nc{\bN}{{\bf N}} \nc{\bO}{{\bf O}}
 \nc{\bP}{{\bf P}} \nc{\bQ}{{\bf Q}} \nc{\bR}{{\bf R}}
 \nc{\bS}{{\bf S}} \nc{\bT}{{\bf T}} \nc{\bU}{{\bf U}}
 \nc{\bV}{{\bf V}} \nc{\bW}{{\bf W}} \nc{\bX}{{\bf X}}
 \nc{\bZ}{{\bf Z}}

\nc{\cA}{{\cal A}} \nc{\cB}{{\cal B}} \nc{\cC}{{\cal C}}
\nc{\cD}{{\cal D}} \nc{\cE}{{\cal E}} \nc{\cF}{{\cal F}}
\nc{\cG}{{\cal G}} \nc{\cH}{{\cal H}} \nc{\cI}{{\cal I}}
\nc{\cJ}{{\cal J}} \nc{\cK}{{\cal K}} \nc{\cL}{{\cal L}}
\nc{\cM}{{\cal M}} \nc{\cN}{{\cal N}} \nc{\cO}{{\cal O}}
\nc{\cP}{{\cal P}} \nc{\cQ}{{\cal Q}} \nc{\cR}{{\cal R}}
\nc{\cS}{{\cal S}} \nc{\cT}{{\cal T}} \nc{\cU}{{\cal U}}
\nc{\cV}{{\cal V}} \nc{\cW}{{\cal W}} \nc{\cX}{{\cal X}}
\nc{\cZ}{{\cal Z}}

\nc{\hA}{{\hat{A}}} \nc{\hB}{{\hat{B}}} \nc{\hC}{{\hat{C}}}
\nc{\hD}{{\hat{D}}} \nc{\hE}{{\hat{E}}} \nc{\hF}{{\hat{F}}}
\nc{\hG}{{\hat{G}}} \nc{\hH}{{\hat{H}}} \nc{\hI}{{\hat{I}}}
\nc{\hJ}{{\hat{J}}} \nc{\hK}{{\hat{K}}} \nc{\hL}{{\hat{L}}}
\nc{\hM}{{\hat{M}}} \nc{\hN}{{\hat{N}}} \nc{\hO}{{\hat{O}}}
\nc{\hP}{{\hat{P}}} \nc{\hR}{{\hat{R}}} \nc{\hS}{{\hat{S}}}
\nc{\hT}{{\hat{T}}} \nc{\hU}{{\hat{U}}} \nc{\hV}{{\hat{V}}}
\nc{\hW}{{\hat{W}}} \nc{\hX}{{\hat{X}}} \nc{\hZ}{{\hat{Z}}}

\def\max{\mathop{\rm max}}

\def\ox{\otimes}

\newcommand{\bra}[1]{\langle#1|}
\newcommand{\ket}[1]{|#1\rangle}
\newcommand{\proj}[1]{| #1\rangle\!\langle #1 |}
\newcommand{\ketbra}[2]{|#1\rangle\!\langle#2|}
\newcommand{\braket}[2]{\langle#1|#2\rangle}

\newcommand{\abs}[1]{|#1|}

\newcommand{\jmp}{J. Math. Phys.~}
\newcommand{\jpa}{J. Phys. A~}

\begin{document}
\title{Length of separable states and symmetrical informationally complete (SIC) POVM }

\author{Lin Chen}
\affiliation{Department of Pure Mathematics and Institute for
Quantum Computing, University of Waterloo, Waterloo, Ontario, N2L
3G1, Canada} \affiliation{Centre for Quantum Technologies, National
University of Singapore, 3 Science Drive 2, Singapore 117542}
\email{cqtcl@nus.edu.sg (Corresponding~Author)}

\begin{abstract}
 This short note reviews the notion and fundamental properties of SIC-POVM and its connection
 with the length of separable states. We also review the t-design.
\end{abstract}

\date{ \today }




\maketitle

\section{definition and background of SIC-POVM}

 \bem

\item
 \cite{zauner99,rbs04,appleby05}
 In the $d$-dimensional Hilbert space, a SIC-POVM consists of $d^2$ outcomes that are subnormalized projectors
 onto pure states $\Pi_j=\frac{1}{d}|\psi_j\rangle\langle\psi_j|$ for
 $j,k=1,\ldots,d^2$, such that
 \begin{eqnarray}\label{eq:SIC}
 |\braket{\psi_j}{\psi_k}|^2=\frac{1+d\delta_{jk}}{d+1}.
 \end{eqnarray}

\item
 \cite[Theorem 2]{rbs04} Using Eq. \eqref{eq:SIC} we can show that
 any SIC-POVM forms a 2-design:
 \bea
 \label{ea:2design}
 \sum^{d^2}_{i=1} \proj{\ps_i,\ps_i}
 =\frac{2d}{d+1}S_d.
 \eea
Here, the operator $S_d$ denotes the $d\times d$ symmetrizer
operator, i.e.,
 \bea
 \label{ea:symmetricOPERATOR}
 S_d := \sum^d_{i=1} \proj{ii}
 +
 \sum^d_{j>i=1} {\ket{ij}+\ket{ji} \over \sqrt2}{\bra{ij}+\bra{ji} \over \sqrt2}.
 \eea

\item
 Eq. \eqref{ea:2design} implies that $\sum^{d^2}_{j=1}\Pi_j=I$, so SIC-POVM is a complete measurement in physics.

\item
 Three basic papers on SIC-POVMs are \cite{zauner99,rbs04,appleby05}.

(1) G. Zauner, "Quantendesigns - Grundz¡§uge einer nicht
kommutativen Designtheorie," PhD thesis (University of Vienna,
1999).

(2) J. M. Renes, R. Blume-Kohout, A. J. Scott, and C. M. Caves, J.
Math. Phys. 45, 2171 (2004). (provide analytical $d=2,3,4$,
numerical $d\le45$.)

(3) D. M. Appleby, J. Math. Phys. 46, 052107 (2005). It provides the
analytical solutions of SIC-POVM for $d=2,\cdots,7,19$.

\item
 \bex
 \label{ex:d=2}
 SIC-POVM for $d=2$. Let
 \bea
 \ket{\ps_0}
 &=&
 \sqrt{3+\sqrt3\over 6} \ket{0}
 +
 e^{\pi i/4}
 \sqrt{3-\sqrt3\over 6}\ket{1}
 ,
 \\
 \ket{\ps_1}
 &=&
 \sqrt{3+\sqrt3\over 6} \ket{0}
 -
 e^{\pi i/4}
 \sqrt{3-\sqrt3\over 6}\ket{1}
 ,
 \\
 \ket{\ps_2}
 &=&
 \sqrt{3+\sqrt3\over 6} \ket{1}
 +
 e^{\pi i/4}
 \sqrt{3-\sqrt3\over 6}\ket{0}
 ,
 \\
 \ket{\ps_3}
 &=&
 \sqrt{3+\sqrt3\over 6} \ket{1}
 -
 e^{\pi i/4}
 \sqrt{3-\sqrt3\over 6}\ket{0}
 ,
 \eea
 . Then one can verify
 \bea
 \sum^3_{i=0} \proj{\ps_i,\ps_i} = \frac{4}{3}S_2.
 \eea
The four states $\ket{\ps_i},i=1,2,3,4$ form a regular tetrahedron
when represented on the Bloch sphere.
 \eex

\item
 Analytical SIC-POVMs have been constructed for dimension $d=2,\cdots,16, 19, 24, 28, 31, 35, 37, 43, 48$, see \cite{sg09}.
 Numerical SIC-POVMs have been constructed for $d\le67$, see the
 details in \cite{zhuTHESIS}. This is achieved by the popular method of Weyl-Heisenberg
 group in quantum information community. However the construction becomes hard for
 higher dimensions. So it is unknown, though widely believed, that whether
 SIC-POVM exists for any dimension $d$.

\item
 Constructing SIC-POVM is one of the most important questions in quantum information. It is related to quantum tomography
 \cite{ze11}, Mutually unbiased bases (MUBs) \cite{wootters06,appleby09}, entanglement theory
 \cite{czw10,zte10}, Lie Algebra \cite{aff11},
 Galois field \cite{aaz12}, foundations of quantum
 mechanics \cite{fs13} and so on.
 \eem

\section{relating SIC-POVM to length}

For a bipartite state $\r$ acting on the Hilbert space $\cH_A \ox
\cH_B$, the partial transpose computed in the standard orthonormal
(o.n.) basis $\{\ket{i}\}$ of system A, is defined by
$\r^\G=\sum_{ij}\ketbra{j}{i}\ox\bra{i}\r\ket{j}$. One can similarly
define the partial transpose $\G_{B}$ on the system $B$. Let $r(\r)$
denote the rank of $\r$. We call the integer pair $(r(\r),r(\r^\G))$
the \textit{birank} of $\r$, and the two integers may be different.
The {\em length}, $L(\r)$, of a separable state $\r$ is the minimal
number of pure product states over all such decompositions of $\r$
\cite{dtt00}. It is known that $L(\r)\ge \max\{r(\r),r(\r^\G)\}$.

One can verify that the partial transpose of the state $\r_2={2\over
d^2+d}S_d$ is
 \bea
 \r_2^\G = {1 \over d^2+d}(I + \proj{\Ps_d}),
 \eea
where $\ket{\Ps_d}=\sum^d_{i=1}\ket{ii}$ is the non-normalized
d-level maximally entangled state. So the separable state $\r_2$ has
birank $({d^2+d \over 2}, d^2)$. Therefore we have $L(\r_2)\ge d^2$.
The equality holds for $d=2$ by Example \ref{ex:d=2}. It also holds
for $d=2,\cdots,16, 19, 24, 28, 31, 35, 37, 43, 48$ \cite{sg09}.
However the question is whether
 \bcj
\label{conj:SICPOVM}
 $L(\r_2)=d^2$ for any $d\ge2$.
 \ecj
The positive answer of this conjecture would imply that the SIC-POVM
exists for any integer $d\ge2$. This argument has been proved by
using the notion of weighted 2-design in \cite[Theorem 4]{scott06}.
On the other hand if Conjecture \ref{conj:SICPOVM} turned out to
fail for some $d$, i.e., $L(\r_2) > d^2$, then SIC-POVM would not
exist for this $d$. This argument has been proved by Eq.
\eqref{ea:2design} and \cite[Theorem 2]{rbs04}.

To conclude, either the positive or negative answer to Conjecture
\ref{conj:SICPOVM} will solve the SIC-POVM problem.

\section{More general background: t-design}

Let $t\ge1$ be an integer. The t-design of dimension d is defined as
a set $S$ of pure product states $\ket{a_i}\in\bC^d$ if
 \bea
 \label{ea:t-design}
 \frac{1}{\abs{S}} \sum_i \proj{a_i}^{\ox t}
 = \r_{t}
 ={d+t-1 \choose t}^{-1} S_{d,t},
 \eea
where $S_{d,t}$ is the $t$-partite symmetrizer operator in the space
$(\bC^d)^{\ox t}$. For example, $S_{d,t}=S_d$ for $t=2$ in Eq.
\eqref{ea:symmetricOPERATOR}. It is known \cite{scott06,bh85} that
the number of design points satisfies
 \bea
 \label{ea:generalLOWERbound}
 \abs{S} \ge
 {d+ \lfloor t/2 \rfloor - 1\choose \lfloor t/2 \rfloor}
 {d+ \lceil t/2 \rceil - 1\choose \lceil t/2 \rceil}.
 \eea
A design which achieves this lower bound is called \textit{tight}.
For example, the bound is equal to $d,d^2$ and $d^2(d+1)/2$ for
$t=1,2,3$, respectively. The t-designs exist for any $d$
\cite{sz84}. In the language of quantum information, it means that
any $t$-partite symmetrizer operator is a non-normalized separable
state. However it is unknown that whether tight t-designs exist,
i.e., whether the length of $t$-partite symmetrizer operator reaches
the lower bound in Eq. \eqref{ea:generalLOWERbound}.

Here are a few known results from the field of t-designs. For $d=2$,
tight t-designs exist for $t=1,2,3,5$ \cite{hs96}. For a few $d>2$,
tight t-designs exist for $t=1,2,3$ \cite{bh85,bh89}. Here is the
detail. It is trivial that tight 1-designs exist for any $d$. The
existence of tight 2-designs is equivalent to the positive answer
for Conjecture \ref{conj:SICPOVM}, in terms of Eq.
\eqref{ea:t-design}. So far this is true for $d=2,\cdots,16, 19, 24,
28, 31, 35, 37, 43, 48$, see \cite{sg09}. Third, the tight 3-designs
are known only for $d = 2, 4, 6$ \cite{hoggar82}. In particular for
$d=2$, the six states from an MUB in $\bC^2$ form a tight 3-design
\cite{zhuTHESIS}. It can also be directly verified by computing the
frame potential.

Note that $\r_t$ is a t-partite separable state. We have
 \bl
\label{le:t-design}
 The tight t-design of dimension $d$ exists if and only if
 $L(\r_t) = {d+ \lfloor t/2 \rfloor - 1\choose \lfloor t/2 \rfloor}
 {d+ \lceil t/2 \rceil - 1\choose \lceil t/2 \rceil}$.
 \el
The proof is based on Ref. [41,42] of \cite{scott06}. Nevertheless,
it is known that the tight t-design does not exist for $d\ge3,t\ge5$
\cite{scott06}.


\section*{Acknowledgments}

I thank Dr. Huangjun Zhu for careful reading this note and pointing
out a few errors in an early version of this note.

\end{document}